\documentclass[12pt]{iopart}
\usepackage{iopams,subfig,psfrag,graphicx}

\newcommand{\sZ}{\mathcal{Z}}
\newcommand{\sF}{\mathcal{F}}

\newcommand{\lep}{\lambda_\varepsilon}
\newcommand{\dbx}{d\bi{x}}
\newcommand{\dby}{d\bi{y}}
\newcommand{\bk}{\bi{k}}

\newcommand{\bq}{\bi{q}}
\newcommand{\bx}{\bi{x}}
\newcommand{\by}{\bi{y}}
\newcommand{\bw}{w}

\begin{document}

\title{Spectral density of random graphs with topological constraints}

\author{Tim Rogers$^1$, Conrad P\'{e}rez Vicente$^2$,  Koujin Takeda$^3$ and Isaac P\'{e}rez Castillo$^1$}
\address{$^1$ Department of Mathematics, King's College London, Strand, London, WC2R 2LS, United Kingdom}
\address{$^2$ Department de F\'{\i}sica Fondamental, Facultat de F\'{\i}sica, Universitat de Barcelona, 08028 Barcelona, Spain}
\address{$^3$ Department of Computational Intelligence and Systems Science, Tokyo Institute of Technology, Yokohama 226-8502, Japan}

\begin{abstract}
The spectral density of random graphs with topological constraints is analysed using the replica method. We consider graph ensembles featuring generalised degree-degree correlations, as well as those with a community structure. In each case an exact solution is found for the spectral density in the form of consistency equations depending on the statistical properties of the graph ensemble in question.  We highlight the effect of these topological constraints on the resulting spectral density.
\end{abstract}

 
\submitto{\JPA}
\maketitle

\section{Introduction}
Since Wigner's seminal work of the 1950's, random matrix theory (RMT) has established itself as a cornerstone of modern theoretical physics, with innumerable applications (see, for example, \cite{Guhr1998} and references therein). One central problem of RMT is the determination of the mean spectral density of an ensemble of random matrices. In some cases, well understood universal laws governing the mean spectral density have been known for some time (see, for instance, \cite{Wigner1958, Dyson1962, Mehta1991}). However, for ensembles of sparse matrices, i.e. matrices with many entries being zero, the picture is rather different. Studied first by Rodgers and Bray \cite{Rodgers1988}, the spectral density of real symmetric sparse random matrices has been extensively researched \cite{Bauer2001,Biroli1999,Dorogovtsev2003,Mirlin1991,Nagao2007,Nagao2008,Semerjian2002}, although exact results have been only obtained relatively recently \cite{Rogers2008,Kuhn2008,Dean2002,Ciliberti2005,Bordenave2007}.\\
One area of research in which sparse random matrices feature heavily is in the modelling of real-world complex networks, with applications in fields as diverse as bioinformatics and  finance. In the investigation of spectral density, the most commonly studied random graph ensembles are those in which the degrees of neighbouring vertices become independent in the limit $N\to\infty$. This class includes the classical Erd\"{o}s-Reyni (or Poissonian) random graphs \cite{Rodgers1988,Bauer2001,Biroli1999,Nagao2007,Nagao2008,Semerjian2002,Dean2002}, and graphs with a specified degree distribution \cite{Kuhn2008,Bordenave2007}. Unfortunately, these simple ensembles may not provide realistic models for real-world complex networks which can include features such as correlations between non-neighbouring vertices, or groups of vertices organised into highly connected communities \cite{10}. Very few results have been obtained for random graph ensembles with more complex topologies, though there exists an approximation scheme for graphs with degree-degree correlations \cite{Dorogovtsev2003} and numerical investigations of some other ensembles \cite{Farkas2001}. \\
In this paper, we  extend the analysis of the spectral density of random  graphs to complex networks with the features described above. In order to achieve this we consider graphs with hierarchically constrained topologies. Introduced in \cite{Nikos,Coolen2008}, such ensembles can be tuned to more closely reflect the statistics of a real graph, whilst remaining in a form for which large $N$ computations are tractable. We also propose a simple generalisation of this ensemble to one featuring a community structure.\\ 
Applying the replica method, expressions are found for spectral densities of these ensembles in terms of the statistical properties of the graphs. For the sake of clarity we only show here the calculations for the spectral density of connectivity matrices, however the techniques can just as well be applied to more general sparse random matrices. This is possible even in the case of non-Hermitian matrices, following a scheme similar to that set out in \cite{Rogers2009}.\\
The ensemble definitions are given in Section 2. In the third section, we prepare for the calculation of the spectral density by first computing relevant statistical properties of the graph ensembles, gaining some insight to the problem on the way. Section 4 contains the replica calculation itself, both for the hierarchically constrained ensemble and for graphs with a community structure. Several applications are discussed in Section 5, where analytical findings of the resulting spectral densities are compared with  numerical diagonalisations. The final section contains a summary and discussion.
\section{Ensemble definitions}
Let $G=(V,E)$ be a graph on a set $V=\{1,...,N\}$ of vertices and $E\subseteq  V\times V$ of edges. The latter set is usually represented by the connectivity matrix $C$, whose entries are $c_{ij}=1$ if $(i,j)\in E$, and $c_{ij}=0$ otherwise $\forall i,j\in V$. We define the $\ell$-th generalised degree of a vertex $i$, denoted as $k_{i}^{(\ell)}(C)$, to be the total number of walks of length $\ell$ starting at vertex $i$. In terms of the connectivity matrix, this is given by the recursive definition
\begin{equation}
k^{(0)}_i(C)=1\,,\quad k^{(\ell)}_i(C)=\sum_{j=1}^Nc_{ij}k^{(\ell-1)}_j(C)\,,\quad\quad \ell=1,\ldots,L\,,
\end{equation}
for some integer $L$. We denote $\bk_i(C)=(k^{(1)}_i(C), \cdots ,k^{(L)}_i(C))\,.$ Note that the first component of $\bk_i(C)$ is simply the degree of vertex $i$, so we will usually refer to it as $k_i(C)$, rather than $k^{(1)}_i(C)$. The generalised degree of a particular vertex contains information about the generalised degrees of its neighbours. For instance, if vertex $i$ has degree $k$ and neighbours $\{j_1,...,j_k\}$, then
\begin{equation}
k^{(\ell)}_i(C)=\sum_{t=1}^{k}k_{j_t}^{(\ell-1)}(C)\quad\textrm{for }\ell=1,...,L\,.
\label{eq:2}
\end{equation}
In what follows we study graph ensembles in which the generalised degrees are constrained. To achieve this, we follow the scheme set out in \cite{Nikos,Coolen2008} and define the weight of the graph ensemble as 
\begin{equation}
\fl W_N(C)=\prod_{i<j}\left[\frac{c}{N}Q(\bk_i,\bk_j)\delta_{c_{ij},1}+\left(1-\frac{c}{N}Q(\bk_i,\bk_j)\right)\delta_{c_{ij},0}\right]\prod_{i=1}^N\delta_{\bk_i(C),\bk_i}\,,
\label{defbW}
\end{equation}
where the $\{\bk_i\}_{i=1}^N$ are taken to be arbitrary, $c$ is the average $c=(1/N)\sum_{i=1}^N k_i$, and $Q(\bk,\bk')$ is a symmetric, non-negative function. Note from   \eref{defbW} that for each vertex $i$, its generalised degree $\bk_{i}(C)$ is constrained to be precisely $\bk_i$. For this reason we also refer to the $\{\bk_i\}_{i=1}^N$ as generalised degrees.\\
As mentioned in the introduction, this ensemble is designed to be amenable to analysis in the large $N$ limit, whilst allowing the topology of the graphs to be tuned by choosing the generalised degrees $\{\bk_i\}$ and the function $Q(\bk,\bk')$. This was demonstrated in \cite{Coolen2008} with the computation of the entropy of the ensemble, and in \cite{Vicente2008} the Ising model is analysed on such graphs in the simpler case of $L=1$.\\
We will also consider an ensemble whose graphs feature community structures. In such graphs, vertices are organised into densely intra-connected clusters, also called modules or communities, with a sparse distribution of inter-community edges \footnote{For a nice and complete review on  community structures and its importance in complex networks see \cite{10}.}. To incorporate this structure without sacrificing the solvability of the model, we propose a generalisation of the previously introduced ensemble \footnote{Very recently, the spectral density of matrices with a particular type of modular form has been studied in \cite{Oraby2007,Jalan2008,Reimer2009}.}.\\
Consider a graph composed of $N$ communities, each of size $M$ (giving a total of $NM$ vertices). We decompose the connectivity matrix $\mathcal{C}$ of this graph into three sets of smaller matrices: a single $N\times N$ connectivity matrix $C$ with entries $c_{ij}=1$ if communities $i$ and $j$ are connected, and zero otherwise; a collection of $M\times M$ matrices $B_{ij}$ encoding the connections between vertices in communities $i$ and $j$; and a collection of $M\times M$ matrices $A_i$ specifying the internal connections of community $i$. For our ensemble, we take the $B_{ij}$ and $A_i$ to be drawn randomly and independently  according to weights $\mu(B)$ and $\nu(A)$, respectively, whilst $C$ is taken from the constrained generalised degree ensemble. All together, the weight for this graph ensemble can be written as follows
\begin{eqnarray}
\fl W^{{\rm com}}_{MN}(\mathcal{C})&=\prod_{i<j}\left[\frac{c}{N}Q(\bk_i,\bk_j)\delta_{c_{ij},1}\mu(B_{ij})+\left(1-\frac{c}{N}Q(\bk_i,\bk_j)\right)\delta_{c_{ij},0}\right]\prod_{i=1}^N\nu(A_i)\delta_{\bk_i(C),\bk_i}
\label{defbWclus}
\end{eqnarray}
 For ease of calculation, we take $\mu$ to satisfy $\mu(B)=\mu(B^T)$. \\
 Exploiting the bridge to statistical mechanics introduced by Edwards and Jones \cite{Edwards1976}, we will compute the mean spectral density of the connectivity matrices of graphs from both ensembles in the limit $N\to\infty$ by the replica method. It will be instructive to first analyse the local statistics of the graphs in the same limit, as this will help us to simplify the subsequent analysis. 
\section{Asymptotic graph statistics}
For the ensembles of graphs under study, the relevant statistical properties are captured by the following joint distribution
\begin{equation}
\fl P\left(\{\bq_t\}_{t=1}^{k},\bk\right)=\lim_{N\to\infty}\left\langle\frac{k}{c N}\sum_{i=1}^N\delta_{\bk_i(C),\bk}\sum_{j_1<\cdots\,<j_k}\delta_{\bk_{j_1}(C),\bq_1}\cdots\delta_{\bk_{j_k}(C),\bq_{k}} c_{ij_1}\cdots c_{ij_{k}}\right\rangle_C\,,
\label{PKN}
\end{equation} 
where here, and hereafter, we use $\langle\cdots\rangle_C$ to denote the ensemble average. Note that \eref{PKN} is the probability of finding a vertex whose neighbour has generalised degree $\bk$ and is connected to $k$ vertices with generalised degrees $\{\bq_t\}_{t=1}^{k}$. Other relevant probability distributions can be obtained by marginalising the expression \eref{PKN}. In particular, by summing with respect to $\{\bq_t\}_{t=1}^{k}$ and denoting the resulting  distribution as $P(\bk)$ we obtain
\begin{equation}
P(\bk)=\lim_{N\to\infty}\left\langle\frac{k}{c N}\sum_{i=1}^N\delta_{\bk,\bk_i(C)}\right\rangle_C\,,
\end{equation}
which is the probability of finding a vertex connected to a vertex with generalised degree $\bk$ (see, for instance, \cite{dorogovtsev2008}). Note that since in our ensemble the generalised degrees are constrained to the arbitrary values $\{\bk_i\}_{i=1}^N$, we have that $P(\bk)=p(\bk) k/c$ with $p(\bk)=\lim_{N\to\infty} (1/N)\sum_{i=1}^N\delta_{\bk,\bk_i}$ being the generalised degree distribution. Other probability distributions, however, may not have such straightforward forms.\\
An expression for the distribution \eref{PKN} can be found through a saddle-point computation, the details of which will prove to be of great use in the later replica analysis of the spectral density, as our final equations for the spectral density will be written in terms of certain marginals of $P\left(\{\bq_t\}_{t=1}^{k},\bk\right)$.
\subsection{General calculation}
To find an expression for $P\left(\{\bq_t\}_{t=1}^{k},\bk\right)$ we need to calculate $\left\langle c_{ij_1}\cdots c_{ij_{k}}\right\rangle_C\,$. To do so we introduce the generating function 
\begin{equation}
Z_N\big(\bi{h}\big)=\sum_CW_N (C)\prod_{i<j}e^{h_{ij}c_{ij}}\,,
\end{equation}
with generating fields $\bi{h}=\{h_{ij}\}$. This allows us to write
\begin{equation}
\left\langle c_{ab_1}\cdots c_{ab_{k}} \right\rangle_C=\frac{1}{Z_N}\left.\frac{\partial^k}{\partial h_{ab_1}\cdots \partial h_{ab_{k}}} Z_N(\bi{h})\right|_{\bi{h}=\textbf{\scriptsize{0}}}\,,
\end{equation}
with $Z_N=Z_N(\textbf{0})$.  To carry out the calculation of $Z_N\big(\bi{h}\big)$ we introduce a Fourier representation for the Kronecker delta constraints appearing in the definition of the weight \eref{defbW}:
\begin{equation}
\delta_{\bk_i(C),\bk_i}=\int_{-\pi}^\pi\frac{d\bi{w}_i}{(2\pi)^L}\exp\left(i\bi{w}_i\cdot\bi{k}_i-i\sum_{\ell=1}^L w_i^{(\ell)}\sum_{j=1}^N c_{ij}k_j^{(\ell-1)} \right)\,,
\end{equation}
with $\bi{w}_i=(w^{(1)}_i,...,w^{(L)}_i)$. The sum over $C$ may now be performed explicitly, obtaining for large $N$
\begin{eqnarray}
\fl\left\langle c_{ab_1}\cdots c_{ab_{k}} \right\rangle_C&=\frac{1}{Z_N}\left(\prod_{t=1}^{k}\frac{c}{N}Q(\bi{k}_a,\bi{k}_{b_t})\right)\int_{-\pi}^\pi\left[\prod_{i=1}^N\frac{d\bi{w}_i}{(2\pi)^L}\right]e^{i\sum_i\bi{w}_i\cdot\bi{k}_i}\nonumber\\
&\times\exp\left[\frac{c}{2N}\sum_{i,j=1}^NQ(\bi{k}_i,\bi{k}_j)\left( e^{-i\sum_{\ell=1}^L\left(w_i^{(\ell)}k_j^{(\ell-1)}+w_j^{(\ell)}k_i^{(\ell-1)}\right)}-1\right)\right]\nonumber\\
&\times\exp\left[-i\sum_{\ell=1}^L\left(w_a^{(\ell)}\sum_{t=1}^{k}k_{b_t}^{(\ell-1)}+k_a^{(\ell-1)}\sum_{t=1}^{k}w_{b_t}^{(\ell)}\right)\right]\,.
\label{CdotsC}
\end{eqnarray}
To integrate out the $w$-variables and apply saddle-point integration we first need to decouple terms comprising vertex indices. To achieve this, we introduce the order parameter
\begin{equation}
\phi(\bk,\bq)=\frac{1}{N}\sum_{i=1}^N\delta_{\bk_i,\bk}\exp\left(-i\sum_{\ell=1}^L\bw_i^{(\ell)} q^{(\ell-1)}\right)\,,
\end{equation}
\newpage
\noindent allowing us to write the average $\left\langle c_{ab_1}\cdots c_{ab_{k}} \right\rangle_C$ as follows
\begin{eqnarray}
\fl\left\langle c_{ab_{1}}\cdots c_{ab_{k}}\right\rangle_C&=\left\langle\frac{\frac{k_a!}{c^{k_a}}\left[\prod_{j=1}^{k_a}\frac{k_{b_j}}{N} Q(\bk_a,\bk_{b_j})\right]\mathbb{I}_{\bk_a}\left(\{\bk_{b_j}\}_{j=1}^{k_a}\right)}{\sum_{\bq_1,\ldots,\bq_{k_a}}\psi(\bk_a,\bq_1)\cdots \psi(\bk_a,\bq_{k_a}) \mathbb{I}_{\bk_a}\left(\{\bq_{j}\}_{j=1}^{k_a}\right)}\right.\nonumber\\
&\hspace{-2cm}\times\left.\prod_{j=1}^{k_a}\frac{\sum_{\bq_1,\ldots,\bq_{k_{b_j}-1}}\psi(\bk_{b_j},\bq_1)\cdots \psi(\bk_{b_j},\bq_{k_{b_j}-1})\mathbb{I}_{\bk_{b_j}}\left(\bk_{a},\{\bq_{j}\}_{j=1}^{k_{b_j}-1}\right)}{ \sum_{\bq_1,\ldots,\bq_{k_{b_j}}}\psi(\bk_{b_j},\bq_1)\cdots \psi(\bk_{b_j},\bq_{k_{b_j}})\mathbb{I}_{\bk_{b_j}}\left(\{\bq_{j}\}_{j=1}^{k_{b_j}}\right) }\right\rangle_{\Phi(\phi,\psi)}\,,
\label{hideous}
\end{eqnarray}
where we have taken  $k=k_a$, and have introduced the indicator function
\begin{equation}
\mathbb{I}_{\bk}\left(\{\bq_t\}_{t=1}^k\right)=\prod_{\ell=1}^L\delta_{k^{(\ell)},\sum_{r=1}^{k} q^{(\ell-1)}_r}\,,
\end{equation}
which enforces the relationship between the generalised degrees of the neighbours of a given vertex, as noted in \eref{eq:2}. The measure in \eref{hideous} is defined by
\begin{equation}
\left\langle\cdots\right\rangle_{\Phi(\phi,\psi)}=\frac{\int\big\{d\phi d\psi\big\} e^{ N\Phi(\phi,\psi)}(\cdots)}{\int\big\{d\phi d\psi\big\}e^{N\Phi(\phi,\psi)}}\,,
\label{measuref}
\end{equation}
with
\begin{eqnarray}
\Phi(\phi,\psi)&=-c\sum_{\bk,\bq}\psi(\bk,\bq)\phi(\bi{k},\bi{q})+\frac{c}{2}\sum_{\bk,\bq}Q(\bi{k},\bq)\phi(\bi{k},\bi{q})\phi(\bi{q},\bi{k})\nonumber\\
&+\sum_{\bi{k}}p(\bi{k})\ln \sum_{\bi{q}_1,\ldots,\bi{q}_k}\psi(\bi{k},\bi{q}_1)\cdots \psi(\bi{k},\bi{q}_k) \mathbb{I}_{\bk}\left(\{\bq_t\}_{t=1}^k\right)\,.
\end{eqnarray}
In the limit $N\to\infty$ this measure converges to a functional Dirac delta centred at the saddle point of $\Phi(\phi,\psi)$. Extremising $\Phi(\phi,\psi)$, we find saddle-point equations
\numparts
\begin{eqnarray}
\hspace{-1cm}\psi (\bk,\bq)=  Q(\bq,\bk)\phi (\bq,\bk)\,,\label{psistar}\\
\hspace{-1cm}\phi (\bk,\bq)= P(\bk)\frac{\sum_{\bq_1,\ldots,\bq_{k-1}} \psi (\bk,\bq_1)\cdots \psi (\bk,\bq_{k-1})\mathbb{I}_{\bk}\left(\bq,\{\bq_t\}_{t=1}^{k-1}\right) }{\sum_{\bq_1,\ldots,\bq_{k}}\psi (\bk,\bq_1)\cdots \psi (\bk,\bq_{k}) \mathbb{I}_{\bk}\left(\{\bq_t\}_{t=1}^{k}\right)}\,.\label{phistar}
\end{eqnarray}
\endnumparts
Finally, returning to the definition of $P\left(\{\bq_t\}_{t=1}^{k},\bk\right)$, using the preceding result together with the saddle-point equations \eref{psistar} and \eref{phistar}, we reach
\begin{eqnarray}
 \fl P\left(\{\bq_t\}_{t=1}^{k},\bk\right)&=\lim_{N\to\infty}\frac{k}{c N}\sum_{i=1}^N\delta_{\bk_i,\bk}\sum_{j_1<\cdots <j_k}\delta_{\bk_{j_1},\bq_1}...\delta_{\bk_{j_k},\bq_{k}}\left\langle  c_{ij_1}\cdots c_{ij_{k}}\right\rangle_C\nonumber\\
\fl&=P(\bk) \frac{ \psi (\bk,\bq_1)\cdots \psi (\bi{k},\bq_k)\mathbb{I}_{\bk}\left(\{\bq_t\}_{t=1}^k\right) }{\sum_{\bi{q}_1,\ldots,\bi{q}_{k}}\psi (\bi{k},\bi{q}_1)\cdots \psi (\bi{k},\bi{q}_{k}) \mathbb{I}_{\bk}\left(\{\bq_t\}_{t=1}^k\right)}\,.
\label{PKNQ}
\end{eqnarray}
It is not our intention to quantify \eref{PKNQ} as this requires the solution of the saddle-point equations, which is a cumbersome task for general values $L$. It is nonetheless interesting to keep in mind this result as it will help us to understand the later results for the spectral density. The following marginals will also be  relevant for our subsequent discussion:
\begin{enumerate}
\item The probability $P(\bq,\bk)$ of finding a pair of connected vertices with generalised degrees $\bk$ and $\bq$:
\begin{eqnarray}
\fl P(\bq,\bk)&=\lim_{N\to\infty}\left\langle\frac{1}{c N}\sum_{i,j=1}^N\delta_{\bk_i(C),\bk}\,\delta_{\bk_{j}(C),\bq}\,  c_{ij}\right\rangle_C\nonumber\\
\fl &=\sum_{\bi{q}_1,\ldots,\bi{q}_{k-1}} P\left(\{\bq_t\}_{t=1}^{k-1},\bq,\bk\right)=Q(\bk,\bq)\phi (\bk,\bq)\phi (\bq,\bk),
\label{Pkk}
\end{eqnarray}
and its conditional distribution
\begin{eqnarray}
 P(\bq|\bk)&=P(\bq,\bk)/P(\bk)=Q(\bk,\bq)\phi (\bk,\bq)\phi (\bq,\bk)\left(\frac{k}{c}p(\bk)\right)^{-1}\,.
\label{Pkkc}
\end{eqnarray}
\item The conditional distribution $P\left(\{\bq_t\}_{t=1}^{k}|\bk\right)$:
\begin{eqnarray}
\fl P\left(\{\bq_t\}_{t=1}^{k}|\bk\right)&= P\left(\{\bq_t\}_{t=1}^{k},\bk\right)/P(\bk)\nonumber\\
&=\frac{ \psi (\bk,\bq_1)\cdots \psi (\bi{k},\bq_k)\mathbb{I}_{\bk}\left(\{\bq_t\}_{t=1}^k\right) }{\sum_{\bi{q}_1,\ldots,\bi{q}_{k}}\psi (\bi{k},\bi{q}_1)\cdots \psi (\bi{k},\bi{q}_{k}) \mathbb{I}_{\bk}\left(\{\bq_t\}_{t=1}^k\right)}\,.
\label{PKNC}
\end{eqnarray} 
\item The conditional distribution $P\left(\{\bq_t\}_{t=1}^{k-1}|\bq,\bk\right)$:
\begin{eqnarray}
\fl P\left(\{\bq_t\}_{t=1}^{k-1}|\bq,\bk\right)=P\left(\{\bq_t\}_{t=1}^{k},\bk\right)/P(\bq,\bk)\nonumber\\
=\frac{\psi (\bi{k},\bi{q}_1) \cdots  \psi (\bi{k},\bi{q}_{k-1})\mathbb{I}_{\bk}\left(\bq,\{\bq_t\}_{t=1}^{k-1}\right)}{\sum_{\bi{q}_1,\ldots,\bi{q}_{k-1}}\psi (\bi{k},\bi{q}_1) \cdots  \psi (\bi{k},\bi{q}_{k-1})\mathbb{I}_{\bk}\left(\bq,\{\bq_t\}_{t=1}^{k-1}\right) }\,.
\label{PKNCQ}
\end{eqnarray} 
\end{enumerate}
\subsection{Case $L=1$}
It is instructive to consider the particular case of $L=1$ as it has also been discussed in previous works (see, for instance, \cite{Coolen2008,Vicente2008}). In this case the generalised degree $\bk$ is simply the degree $k$. As we only constrain the degrees, the indicator functions in \eref{phistar} are identically one, giving
\begin{equation}
\phi (k,q)=\frac{P(k)}{\sum_{q} \psi (k,q) }\,,\quad\quad \psi (k,q)=Q(k,q)\phi (q)\,.
\end{equation}
with $P(k)=p(k)k/c$ being the degree distribution of the nearest neighbour of a vertex. Since the right hand side of the first preceding equation is independent of $q$, so is $\phi (k,q)$, and therefore we write $\phi (k,q)=\phi (k)$. Besides, introducing $\psi (k)=\sum_{q}\psi (k,q)$, we reach
\begin{equation}
\phi (k)=\frac{P(k)}{\psi (k)}\,,\quad\textrm{and}\quad\psi (k)=\sum_{q}Q(k,q)\phi (q)\,.
\label{phipsistar}
\end{equation}
From this, we obtain simplified expressions for   \eref{Pkk} and \eref{Pkkc}:
\numparts
\begin{eqnarray}
P(q,k)&=P(k)P(q)\frac{Q(k,q)}{\psi (k)\psi (q)}=P(k)\frac{\psi (k,q)}{\psi (k)}\,,\label{PkkQ}\\
P(q|k)&=P(q)\frac{Q(k,q)}{\psi (k)\psi (q)}=\frac{\psi (k,q)}{\psi (k)}\,.\label{PkkcQ}
\end{eqnarray}
\endnumparts
Moreover, this also induces factorisation in \eref{PKNC} and \eref{PKNCQ}, giving
\begin{equation}
\fl P\left(\{q_t\}_{t=1}^{k}|k\right)=\prod_{t=1}^{k}P(q_t|k)\,,\quad\textrm{and}\quad P\left(\{q_t\}_{t=1}^{k-1}|q,k\right)=P\left(\{q_t\}_{t=1}^{k-1}|k\right)\,.
\end{equation}
Suppose further that we have a separable function $Q(k,q)=v(k)v(q)$. Then, from  \eref{phipsistar} is  easy to see that $\sum_{q}v(q)\phi (q)=\pm 1$,  giving the solution $\psi (k)=\pm v(k)$. This implies in turn that $P(q|k)=P(q)$ as expected.\\
Having gained some understanding of the typical order parameters involved in the problem, we move on to the explicit calculation of the spectral density. 
\section{The spectral density of constrained graphs}
Suppose $C$ is the connectivity matrix of a random graph belonging to the ensemble under consideration. It is real and symmetric and hence has $N$ real eigenvalues, which we denote by $\big\{\lambda^C_{i}\big\}_{i=1}^N$.  The natural object of study is the mean spectral density in the limit $N\to\infty$,
\begin{equation}
\rho(\lambda)=\lim_{N\to\infty}\left\langle\frac{1}{N}\sum_{i=1}^N\delta\left(\lambda-\lambda^C_{i}\right)\right\rangle_C\,.
\label{density}
\end{equation}
To compute \eref{density}, we first recast the problem in terms of a Gaussian integral. Following Edwards and Jones \cite{Edwards1976}, one may write
\begin{equation}
\rho(\lambda)=-\lim_{\varepsilon\rightarrow 0^+}\lim_{N\to\infty}\frac{2}{\pi N}\textrm{ Im }\frac{\partial}{\partial  \lambda }\left\langle\ln\sZ_{C}(\lep)\right\rangle_C\,,
\label{avlog}
\end{equation}
where $\lep=\lambda-i\varepsilon$ and $\sZ_C$ is given by
\begin{equation}
 \sZ_{C}(\lep)=\int\left[\prod_{i=1}^Ndx_i\right]\exp\left(-i\frac{\lep}{2}\sum_{i=1}^N x_i^2+i\sum_{i<j} c_{ij}x_ix_j\right)\,.
\label{partition}
\end{equation}
Ignoring the imaginary units, this object is reminiscent of the partition function of a system of dynamical variables interacting on a graph, much like the Gaussian ferromagnetic model introduced in \cite{Berlin1952}. \\
Reasoning along these lines, the tools of statistical mechanics can be brought to bear on the calculation of spectral density. In particular, the replica method has been frequently applied, leading either to approximative schemes \cite{Biroli1999,Nagao2007,Nagao2008,Semerjian2002}, or more recent exact solutions \cite{Kuhn2008,Dean2002,Reimer2009}. We proceed with the analysis for our ensemble, paying close attention to the impact of constraining the generalised degrees.
\subsection{General calculation}
To evaluate the average of the logarithm in \eref{avlog}, we apply the replica method, writing
\begin{equation}
\left\langle\ln\sZ_{C}(\lep)\right\rangle_C=\lim_{n\to0}\frac{1}{n}\ln\left\langle\sZ^n_{C}(\lep)\right\rangle_C\,,
\label{logav}
\end{equation}
where the replicated partition function reads
\begin{equation}
\fl\left\langle\sZ^n_{C}(\lep)\right\rangle_C=\int\left[\prod_{i=1}^Nd\bx_i\right]\exp\left(-i\frac{\lep}{2}\sum_{i=1}^N \bi{x}_i^2\right)\left\langle\exp\left(i\sum_{i<j} c_{ij}\bi{x}_i\cdot\bi{x}_j\right)\right\rangle_C\,,
\end{equation}
with $\bi{x}_i=(x_i^{(1)}, \ldots,x_i^{(n)})$ and $d \bx_i=dx_i^{(1)}\cdots dx_i^{(n)}$. As in the previous calculation, we use a Fourier representation for the Kronecker deltas in the weight \eref{defbW} to enable us to perform the ensemble average explicitly, writing, for large $N$,
\begin{eqnarray}
&\fl\left\langle\sZ^n_C(\lep)\right\rangle_{C}=\frac{1}{Z_N}\int\left[\prod_{i=1}^Nd\bx_i\right]\int_{-\pi}^\pi\left[\prod_{i=1}^N\frac{d\bi{w}_i}{(2\pi)^L}\right]e^{i\sum_{i=1}^N\bi{w}_i\cdot\bi{k}_i-\frac{i \lep}{2}\sum_{i=1}^N\bi{x}^2_i}\nonumber\\
&\fl\quad\times\exp\left[\frac{c}{2N}\sum_{i,j=1}^N Q(\bi{k}_i,\bi{k}_j)\left(e^{-i\sum_{\ell=1}^L\left(w_i^{(\ell)}k_j^{(\ell-1)}+w_j^{(\ell)}k_i^{(\ell-1)}\right)+i\bi{x}_i\cdot\bi{x}_j}-1\right) \right]\,.
\end{eqnarray}
To decouple vertices we introduce the following  order parameter
\begin{equation}
\Phi(\bi{x},\bk,\bq)=\frac{1}{N}\sum_{i=1}^N\delta_{\bk_i,\bk}\delta(\bi{x}_i-\bi{x})\exp\left(-i\sum_{\ell=1}^Lw_i^{(\ell)}q^{(\ell-1)}\right)\,,
\end{equation}
with $\bi{x}=(x^{(1)},\ldots, x^{(n)})$ and $\delta(\bi{x}_i-\bi{x})=\delta(x^{(1)}_i-x^{(1)})\cdots\delta(x^{(n)}_i-x^{(n)})$. After enforcing the order parameter using a functional Dirac delta and rearranging terms we can write a compact expression for the replicated partition function:
\begin{equation}
\left\langle\sZ^n_C(\lep)\right\rangle_{C}=\int\{d\Phi d\Psi\} e^{N\sF(\Phi,\Psi)}\,,
\label{sZsF}
\end{equation}
where 
\begin{eqnarray}
\fl\sF(\Phi,\Psi)&=-c\sum_{\bk,\bq}\int d\bx\Psi(\bx,\bk,\bq)\Phi(\bx,\bk,\bq)-\frac{c}{2}\sum_{\bk,\bq}p(\bk)p(\bq)Q(\bk,\bq)\nonumber\\
\fl&+\frac{c}{2}\sum_{\bk,\bq}Q(\bk,\bq)\int\dbx\dby\Phi(\bx,\bk,\bq)\Phi(\by,\bq,\bk)e^{i\bi{x}\cdot\bi{y}}\\
\fl&+\sum_{\bi{k}}p(\bi{k})\ln\frac{c^{k}}{k!}\int d\bi{x}e^{-i\frac{\lep}{2}\bi{x}^2}\sum_{\bq_1,\ldots,\bq_{k}}\Psi(\bx,\bk,\bq_1)\cdots\Psi(\bx,\bk,\bq_{k})\mathbb{I}_{\bk}\left(\{\bq_t\}_{t=1}^{k}\right)\,.\nonumber
\end{eqnarray}
The integral \eref{sZsF} can now be evaluated by steepest descent. Extremising $\sF$ with respect to $\Phi$ and $\Psi$ we obtain the saddle-point equations
\numparts
\begin{eqnarray}
\fl\Psi (\bx,\bi{k},\bi{q})=Q(\bi{k},\bi{q})\int\dby\Phi (\by,\bi{q},\bi{k})e^{i\bx\cdot\by}\,,
\label{Psistar}\\
\fl\Phi (\bx,\bi{k},\bi{q})=P(\bk)\frac{e^{-i\frac{\lep}{2}\bx^2}\sum_{\bi{q}_1,\ldots,\bi{q}_{k-1}}\Psi (\bx,\bi{k},\bi{q}_1)\cdots\Psi (\bx,\bi{k},\bi{q}_{k-1})\mathbb{I}_{\bk}\left(\bq,\{\bq_t\}_{t=1}^{k-1}\right)}{\int\dby e^{-i\frac{\lep}{2}\by^2}\sum_{\bi{q}_1,\ldots,\bi{q}_{k}}\Psi (\by,\bi{k},\bi{q}_1)\cdots\Psi (\by,\bi{k},\bi{q}_k)\mathbb{I}_{\bk}\left(\{\bq_t\}_{t=1}^{k}\right)}\,.
\label{Phistar}
\end{eqnarray}
\endnumparts
The natural next step in the calculation is to make a replica symmetric ansatz. In the case of graphs with unconstrained topologies, the correct form for the order parameters has been recently established as a superposition of Gaussians \cite{Rogers2008,Kuhn2008, Dean2002}. With a modest amount of foresight we extend this to the correlated case by writing
\numparts
\begin{eqnarray}
  \Phi (\bi{x},\bk,\bq)=\phi(\bk,\bq)\int d\Delta\,\phi(\Delta|\bk,\bq)\prod_{\alpha=1}^n \frac{e^{-\frac{1}{2\Delta}(x^{(\alpha)})^2}}{\sqrt{2\pi \Delta}}\,,\\
\Psi (\bi{x},\bk,\bq)=\psi(\bk,\bq)\int d\Delta\,\psi(\Delta|\bk,\bq)\prod_{\alpha=1}^n \frac{e^{-\frac{\Delta}{2}(x^{(\alpha)})^2}}{\sqrt{2\pi/ \Delta}}\,.
\end{eqnarray}
\endnumparts
where we assume that the densities $\psi(\Delta|\bk,\bq)$ and $\phi(\Delta|\bk,\bq)$ are  normalised, i.e. $\int d\Delta\,\psi(\Delta|\bk,\bq)=1$ and similarly for $\phi(\Delta|\bk,\bq)$. Note that the parameter $\Delta$ is generally a complex variable and $d\Delta= d{\rm Re}\Delta\, d{\rm Im}\Delta$. Plugging the ans\"atze into the saddle-point equations and taking the replica limit $n\to 0$ we obtain
\numparts
\begin{eqnarray}
\fl\psi(\Delta|\bk,\bq)\psi(\bk,\bq)=Q(\bi{k},\bi{q})\phi(\Delta|\bi{q},\bi{k})\phi(\bi{q},\bi{k})\,,\label{tau1}\\
\fl\phi(\Delta|\bi{k},\bi{q})=\sum_{\bi{q}_1,\ldots,\bi{q}_{k-1}}\left\{\frac{P(\bi{k})}{\phi(\bi{k},\bi{q})}\frac{\psi(\bi{k},\bi{q}_1)\cdots\psi(\bi{k},\bi{q}_{k-1})\mathbb{I}_{\bk}\left(\bq,\{\bq_t\}_{t=1}^{k-1}\right) }{\sum_{\bi{m}_1,\ldots,\bi{m}_{k}} \psi(\bi{k},\bi{m}_1)\cdots\psi(\bi{k},\bi{m}_k)\mathbb{I}_{\bk}\left(\{\bi{m}_t\}_{t=1}^{k}\right)}\right\}\nonumber\\
\times\int\left[\prod_{t=1}^{k-1} d\Delta_t\,\psi(\Delta_t|\bi{k},\bi{q}_t)\right]\delta\left(\Delta-\frac{1}{i\lep+\sum_{t=1}^{k-1}\Delta_t}\right)\,.
\label{pi1}
\end{eqnarray}
\endnumparts
Integrating with respect to $\Delta$ in \eref{tau1} and \eref{pi1} reveals
\numparts
\begin{eqnarray}
 \psi(\bk,\bq)=Q(\bi{k},\bi{q})\phi(\bi{q},\bi{k})\,,\\
 \phi(\bi{k},\bi{q})=P(\bi{k})\frac{\sum_{\bi{q}_1,\ldots,\bi{q}_{k-1}} \psi(\bi{k},\bi{q}_1)\cdots\psi(\bi{k},\bi{q}_{k-1})\mathbb{I}_{\bk}\left(\bq,\{\bq_t\}_{t=1}^{k-1}\right)}{\sum_{\bi{m}_1,\ldots,\bi{m}_{k}} \psi(\bi{k},\bi{m}_1)\cdots\psi(\bi{k},\bi{m}_k)\mathbb{I}_{\bk}\left(\{\bi{m}_t\}_{t=1}^{k}\right)}\,,
\end{eqnarray}
\endnumparts
and we recognise these equations as precisely the saddle-point equations \eref{psistar} and \eref{phistar} in the earlier calculation of the asymptotic graph statistics. This fact is not surprising since these objects depend only upon the generalised degrees $(\bk,\bq)$, and in this regard the order parameters in both calculations are the same. In turn, this also reveals that the expression between braces in   \eref{pi1} is precisely the conditional distribution $P\left(\{\bi{q}_t\}_{t=1}^{k-1}|\bq,\bi{k}\right)$. All in all, we can write a self-consistency equation for the density $\psi(\Delta|\bq,\bk)$:
\begin{eqnarray}
\psi(\Delta|\bi{q},\bi{k})&=\sum_{\bi{q}_1,\ldots,\bi{q}_{k-1}}P\left(\{\bi{q}_t\}_{t=1}^{k-1}|\bq,\bi{k}\right)\nonumber\\
&\times \int\left[\prod_{t=1}^{k-1} d\Delta_t\,\psi(\Delta_t|\bi{k},\bi{q}_t)\right]\delta\left(\Delta-\frac{1}{i\lep+\sum_{t=1}^{k-1}\Delta_t}\right) \,.
\label{taufinal}
\end{eqnarray}
The simple form of this equation is possible thanks to the link between the order parameters in this calculation and the asymptotic form of the conditional distribution \eref{PKNCQ}. Moreover, written in this way, the physical interpretation is clear: $\psi(\Delta|\bq,\bk)$ is the conditional density of the parameter  $\Delta$ for a vertex of generalised degree $\bk$, given that it has a neighbour of generalised degree $\bq$.\\
All that remains now is to compute the spectral density.  From   \eref{avlog} we can write
\begin{eqnarray}
\rho(\lambda)&=-\lim_{\varepsilon\rightarrow 0^+}\lim_{n\to0}\frac{2}{\pi n }\textrm{ Im }\frac{\partial}{\partial  \lambda }\sF(\Phi ,\Psi )\nonumber\\
&=\lim_{\varepsilon\to0^{+}}\frac{1}{\pi}\textrm{Re}\sum_{\bk}p(\bk)\int d\Delta \, \psi_{\rm{phys}}(\Delta|\bk)\, \Delta\,,
\label{rhofinal}
\end{eqnarray}
with
\begin{equation}
\fl\psi_{\rm{phys}}(\Delta|\bk)=\hspace{-3pt}\sum_{\bi{q}_{1},\ldots,\bi{q}_{k}}\hspace{-2pt}P(\{\bi{q}_t\}_{t=1}^{k}|\bi{k})\int\left[\prod_{t=1}^{k} d\Delta_t\,\psi(\Delta_t|\bk,\bq_t)\right]\delta\left(\Delta-\frac{1}{i\lep+\sum_{t=1}^k\Delta_t}\right).
\label{tauphys}
\end{equation}
The calculation is now complete. We have expressed the limiting mean spectral density for the connectivity matrices of graphs from the ensemble defined by \eref{defbW} in terms of densities $\psi(\Delta|\bq,\bk)$, moreover, we have found that the consistency equation \eref{taufinal} for these densities is phrased simply in terms of the asymptotic generalised degree statistics of the graphs.\\
Although, as this calculation has shown, the spectral density of the ensemble studied here is governed entirely by its generalised degree statistics, we should point out that this is by no means the general rule.
\subsection{Case $L=1$}
Let us again consider the particular case $L=1$. As before the generalised degree $\bk$ is simply the degree $k$, and we have $P\left(\{q_t\}_{t=1}^{k-1}|q,k\right)=P\left(\{q_t\}_{t=1}^{k-1}|k\right)$. This implies that the left hand side of   \eref{taufinal} does not depend on $q$, that is $\psi(\Delta|q,k)=\psi(\Delta|k)$. This implies in turn on the right hand side of \eref{taufinal} that $\psi(\Delta_t|k,q_t)=\psi(\Delta_t|q_t)$. Moreover, since $P\left(\{q_t\}_{t=1}^{k-1}|k\right)=\prod_{t=1}^{k-1}P(q_t|k)$ and upon defining $\varphi(\Delta|k)=\sum_{q}P(q|k)\psi(\Delta|q)$, we can write the following self-consistency equation for $\varphi(\Delta|k)$:
\begin{equation}
\fl\quad\quad \varphi(\Delta|k)=\sum_{q}P(q|k)\int\left[\prod_{t=1}^{q-1} d\Delta_t\, \varphi(\Delta_t|q)\right]\delta\left(\Delta-\frac{1}{i\lep+\sum_{t=1}^{q-1}\Delta_t}\right)\,.
\label{taucorr}
\end{equation}
The spectral density is given by
\begin{equation}
\rho(\lambda)=\lim_{\varepsilon\to0^{+}}\frac{1}{\pi}\textrm{Re}\sum_{k}p(k)\int d\Delta\,  \psi_{\rm{phys}}(\Delta|k)\, \Delta\,,
\label{rhocorr}
\end{equation}
with
\begin{equation}
\psi_{\rm{phys}}(\Delta|k)=\int\left[\prod_{t=1}^{k} d\Delta_t\, \varphi(\Delta_t|k)\right]\delta\left(\Delta-\frac{1}{i\lep+\sum_{t=1}^k\Delta_t}\right)\,.
\label{physcorr}
\end{equation}
The situation is further simplified if $Q$ is taken to be separable, in which case $P(q|k)=P(q)$ and we recover the results of \cite{Kuhn2008} for the uncorrelated case.
\subsection{Graphs with community structure}
We turn our attention to graphs with communities whose ensemble weight is given by \eref{defbWclus}. In this situation it is again possible to compute expressions for the mean spectral density of the ensemble using the replica method. Let us briefly outline the main features of the calculation.\\ 
Introducing $N$ vectors $\{X_i\}_{i=1}^N$, of $M$ components each, $X_i=(x_{i,1},\ldots,x_{i,M})$, one may write
\begin{equation}
\rho(\lambda)=-\lim_{\varepsilon\rightarrow 0^+}\lim_{N\to\infty}\frac{2}{\pi NM}\textrm{ Im }\frac{\partial}{\partial  \lambda }\Big\langle\ln\sZ_\mathcal{C}(\lep)\Big\rangle_{\mathcal{C}}\,,
\end{equation}
where $\langle\cdots\rangle_\mathcal{C}$ denotes the ensemble average and 
\begin{equation}
 \fl\sZ_\mathcal{C}(\lep)=\int\left[\prod_{i=1}^N dX_i\right]\exp\left(-\frac{i}{2}\sum_{i=1}^N X_i\left(\lep I_M-A_i\right) X_i^T+i\sum_{i<j} c_{ij}X_iB_{ij}X_j^T\right)\,,
\label{partitionComm}
\end{equation}
and $dX_i=\prod_{m=1}^M dx_{i,m}$. We consider the thermodynamic limit to be given by $N\to\infty$, whilst $M$ remains fixed and finite. Treating $X_i$  as individual vector-valued dynamical variables the calculation proceeds as usual via the replica method. The order parameters take the same form as in the previous calculation, however  with the replica symmetric ansatz parameterised by $M\times M$ matrices $\bi{\Delta}$. In the end, one obtains the following self-consistency equation:
\begin{eqnarray}
&\fl\psi(\bi{\Delta}|\bq,\bk)=\sum_{\bq_1...\bq_{k-1}}P(\{\bq_t\}_{t=1}^{k-1}|\bq,\bk)\int\left[\prod_{t=1}^{k-1}d\bi{\Delta}_t\psi(\bi{\Delta}_t|\bk,\bq_t,)\,dB_t\mu(B_t)\right]\nonumber\\ 
&\qquad\qquad\int dA\nu(A) \delta\left(\bi{\Delta}-\left(i(\lep I_M-A)+\sum_{t=1}^{k-1}B_t\bi{\Delta}_tB^T_t\right)^{-1}\right)\,.\label{taucomm}
\end{eqnarray}
The spectral density is recovered via
\begin{eqnarray}
\rho(\lambda)&=\lim_{\varepsilon\rightarrow 0^+}\frac{1}{\pi M }{\rm Re }\sum_{\bi{k}}p(\bi{k})\int d\bi{\Delta}\,\psi_{{\rm phys}}(\bi{\Delta}|\bi{k}){\rm Tr}\bi{\Delta}
\label{rhocomm}
\end{eqnarray}
where
\begin{eqnarray}
&\fl\psi_{{\rm phys}}(\bi{\Delta}|\bi{k})= \sum_{\bq_1...\bq_{k}}P(\{\bq_t\}_{t=1}^{k}|\bk)\int\left[\prod_{t=1}^{k}d\bi{\Delta}_t\psi(\bi{\Delta}_t|\bk,\bq_t)\,dB_t\mu(B_t)\right]\nonumber\\ 
&\qquad\qquad\int dA\nu(A) \delta\left(\bi{\Delta}-\left(i(\lep I_M-A)+\sum_{t=1}^{k}B_t\bi{\Delta}_tB^T_t\right)^{-1}\right)\,.
\end{eqnarray}
Note that in the case $M=1$ (that is, when each community is a single vertex) we have $\nu(A)=\delta_{A,0}$, and $\mu(B)=\delta_{B,1}$ and we recover the result of the previous calculation. Also, for general $M$, taking $L=1$ induces the same simplifications as observed previously.
\section{Numerics}
In general the self-consistency equation \eref{taufinal} is not exactly solvable. However a numerical solution may be efficiently obtained using population dynamics \cite{Mezard2001,Kuhn2008}. Although we will consider mainly the case $L=1$, we describe this numerical procedure for the general case. First, for all possible pairs of generalised degrees $(\bq,\bk)$, each density $\psi(\Delta|\bq,\bk)$ in   \eref{taufinal} is represented by a population of $\mathcal{N}$ variables $\{\Delta_i(\bq,\bk)\}_{i=1}^{\mathcal{N}}$. The following procedure is then repeated a predefined number of iteration steps:
\begin{enumerate}
\item[1.] Choose a pair of generalised degrees $(\bq,\bk)$ and a variable $\Delta_a(\bq,\bk)$ uniformly at random from its population.
\item[2.] Randomly select set of generalised degrees $\{\bq\}_{t=1}^{k-1}$ according to the distribution $P\left(\{\bi{q}_t\}_{t=1}^{k-1}|\bq,\bi{k}\right)$.
\item[3.] Choose $k-1$ variables $\{\Delta_{\ell_1}(\bk,\bq_1),\ldots,\Delta_{\ell_{k-1}}(\bk,\bq_{k-1})\}$ uniformly at random from their populations.
\item[4.] Assign $(i\lep+\sum_{t=1}^{k-1}\Delta_{\ell_t}(\bk,\bq_t))^{-1}\to \Delta_a(\bq,\bk)$
\end{enumerate}
This procedure is adapted straightforwardly to find a numerical solution of $\psi_{\rm{phys}}(\Delta|\bk)$ using   \eref{tauphys}, which is then used to calculate the spectral density from   \eref{rhofinal}. \\
To assess our results we compare our analytical findings with results from numerical diagonalisation of graphs. Although the starting point of our calculations was the ensemble definition \eref{defbW}, the final equations are phrased only in terms of the resulting generalised degree statistics. It is therefore appropriate to compare our results directly to data coming from random graphs with prescribed generalised degree statistics, without concerning ourselves with intermediate step of determining a suitable choice of weight $W_N(C)$.  For $L=1$  we heuristically adapt the Steger and Wormald algorithm \cite{Steger1999} \footnote{We have also tried an adapted version of the algorithm suggested in \cite{1459763} with similar results.} to generate graphs with a given a connected degree-degree distribution $P(k,k')$ in the following way: given a degree sequence  $\overline{k}=(k_1,\ldots,k_N)$ with  number of edges $m=\frac{1}{2}\sum_{i=1}^N k_i$, iterate the following procedure:
\begin{enumerate}
\item[1.] Let $E$ be a set of assigned edges, $\widehat{k}=(\widehat{k}_1,\ldots,\widehat{k}_N)$ an $N$-tuple of integers.
\item[2.] Initialise $E=\emptyset$, $\widehat{k}=\overline{k}$
\item[3.] Choose two vertices $v_i,v_j\in V$ with probability $p_{ij}\propto P(k_i,k_j)\widehat{k}_i\widehat{k}_j$ and $(v_i,v_j)\not\in E$. Reduce $\widehat{k}_i,\widehat{k}_j$ by 1.
\item[4.] Repeat Step 3 until no more edges can be added to $E$.
\item[5.] If $|E|<m$ report failure otherwise output graph. 
\end{enumerate}
The input for this algorithm is the degree sequence of the graphs to be generated, however, our results are expressed in terms of the degree distribution $p(k)$. We therefore  need to generate degree sequences which are compatible with $p(k)$. We discuss two possibilities:
\begin{description}
\item[Random degree sequence:] For each instance, the degrees are randomly drawn from $p(k)$. There is a chance that no graph can be generated exactly fitting the resulting degree sequence $\overline{k}$. In this case the degree sequence is said to be non-graphical. To deal with this, one may choose either to check the graphicality of $\overline{k}$ before generating graphs, or simply accept all graphs generated regardless of whether they fail step 5. There are various ways to check graphicality, for instance, a theorem of Erd\"{o}s and Gallai states that a degree sequence $\overline{k}$ with $k_1\geq \cdots\geq k_N$ and $\sum_{i=1}^Nk_i$ even is graphical if and only if, for all $n=1,...,N-1$
\begin{equation}
\sum_{i=1}^nk_i\leq n(n-1)+\sum_{i=n+1}^N\min\{k_i,n\}\,.
\end{equation}
For a discussion on graphicality and the generation of random graphs, see \cite{Blitzstein2006}.
\item[Fixed degree sequence:] Select a set of positive integers $\{N,N_1,N_2,\ldots\}$ such that $N=\sum_{k}N_k$ and $p(k)\simeq N_k/N$. We then generate random degree sequences with $N_1$ vertices of degree one, $N_2$ vertices of degree 2, and so on.  
 \end{description}
In our experience, the adapted Steger-Wormald algorithm yields graphs with the desired properties and produces almost no failures, provided one selects the appropriate method of generating degree sequences.
\subsection{Case $L=1$. Correlated degrees}
\begin{figure}
\psfrag{y}[B][B][1][-90]{$\rho(\lambda)\quad$}\psfrag{x}{$\lambda$}
\includegraphics[trim=0 50 0 50, width=\textwidth]{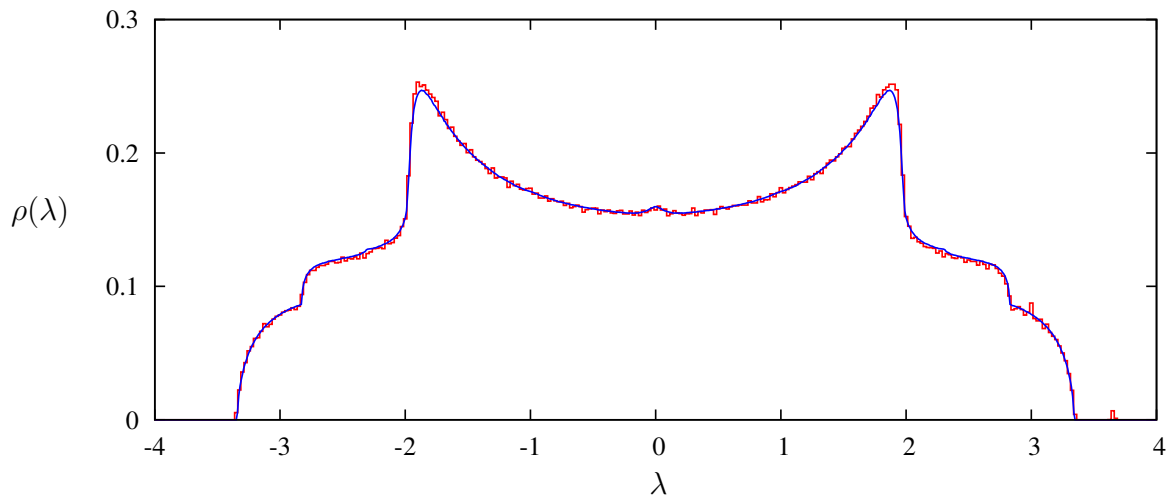}
\caption{Comparison of the results of population dynamics (blue line) and direct diagonalisation (red histogram) for the choice of $P(k'|k)$ given in \eref{arbP}. To construct the histogram we use the adapted Steger-Wormald algorithm to generate 1000 graphs of size $N=2000$. The degree sequences were generated randomly according to $p(k)$ given in \eref{arbP} and only eight failures were reported.}
\label{arb}
\end{figure}

\begin{figure}
\psfrag{y}[B][B][1][-90]{$\rho(\lambda)\quad$}\psfrag{x}{$\lambda$}
\psfrag{k20}[B][B][0.8][0]{$k=20^2$}\psfrag{k18}[B][B][0.8][0]{$k=18^2$}\psfrag{k16}[B][B][0.8][0]{$k=16^2$}\psfrag{k14}[B][B][0.8][0]{$k=14^2$}
\includegraphics[trim=0 50 0 50, width=\textwidth]{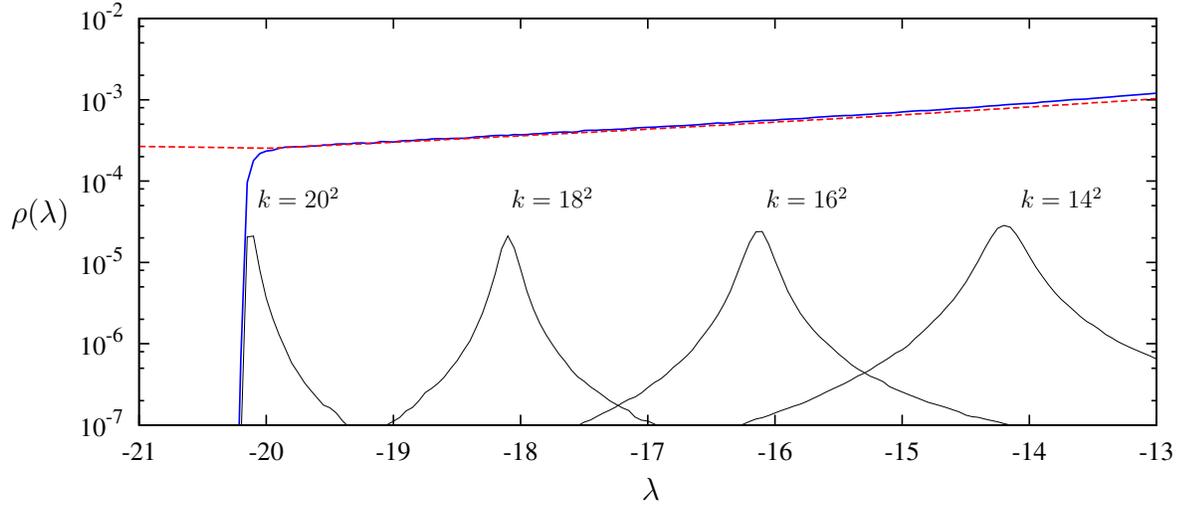}
\caption{Detail of the tail of the spectral density for the choice $P(k,k')\propto\tau^{kk'}$, where $\tau=0.999$ and $k_{\max}=400$. The continuous blue line is the full result from population dynamics, with the labelled black lines being isolated contributions from vertices of high degree. The dashed red line shows an estimate for the tail derived from the effective medium approximation.}
\label{tail}
\end{figure}

Let us consider a graph ensemble in which the degrees of neighbouring vertices are correlated, that is, $P(k,k')$ does not factorise. For the sake of simplicity, we consider graphs whose vertices can only have degrees 2, 3 and 4, with the following degree distribution $p(k)$ and conditional distribution $P(k'|k)$:
\begin{eqnarray}
p(k)=\frac{18}{37}\delta_{k,2}+\frac{4}{37}\delta_{k,3}+\frac{15}{37}\delta_{k,4}\,,\nonumber\\
P(k'|k)=\frac{2}{3}\delta_{k,2}\delta_{k',2}+\delta_{k,3}\delta_{k',3}+\frac{4}{5}\delta_{k,4}\delta_{k',4}+\frac{1}{5}\delta_{k,4}\delta_{k',2}+\frac{1}{3}\delta_{k,2}\delta_{k',4}\,.
\label{arbP}
\end{eqnarray}
To compute the spectral density of the resulting ensemble, we solve the self-consistency equation \eref{taucorr} using population dynamics as described previously. In this particular case, for each value of $k\in\{2,3,4\}$ the corresponding density $\varphi(\Delta|k)$ is represented by a population of $\mathcal{N}=10^4$  variables $\{\Delta_{i}(k)\}_{i=1}^{\mathcal{N}}$, which are iterated over 200 MC steps. The spectral density is then computed via   \eref{physcorr} and \eref{rhocorr}. To obtain smooth results, the spectral density is averaged over a further 50 MC steps. \\
For comparison,  we have also calculated the spectral density by numerically diagonalising 1000 graphs of size $N=2000$. In this case, each instance is produced by first generating a degree sequence $\overline{k}=(k_1,\ldots,k_N)$ according to $p(k)$ and then applying the adapted Steger-Wormald algorithm as described previously.  We have checked that both methods of generating the degree sequences produce equivalent results for large samples and that the number of failures of the algorithm is negligible compared to the sample size.
\\
The results of population dynamics and numerical diagonalisation are presented in Figure \ref{arb} which, apart from peaks at $\lambda=3$ and $\lambda\simeq 3.7$ due to finite size effects, shows excellent agreement.\\
Note from the choice of $P(k'|k)$ that there is a bias towards edges between vertices of the same degree, which suggests that the spectral density should share some features with the spectral densities of regular graphs of degrees 2, 3 and 4. Indeed, there are peaks close to $\pm2$ and $\pm\sqrt{8}$, coming from peaks in the spectral density of regular graphs of degree 2 and 3, and the domain of the spectral density is approximately $\big[-2\sqrt{3},2\sqrt{3}\big]$, the domain of the spectral density of the regular graph of degree 4.
\\
We consider next an ensemble with a power-law degree distribution and correlated degrees. Such models arise often in the study of real world complex networks. We have chosen $P(k,k')\propto\tau^{kk'}$, where $\tau<1$, with a maximum value for $k$ being $k_{{\rm max}}$. In the limits $k_{\max}\to\infty$ and $\tau\to1$, this choice results in a power-law degree distribution with exponent 2, though for the purpose of simulations, we will take $\tau=0.999$, and keep $k_{\max}$ finite.\\ 
We use the previously explained algorithm to generate  graphs with distribution $P(k,k')$. For graphs of size $N=2000$ it is necessary to take a rather low maximum degree $k_{{\rm max}}=45 \,(\,\simeq\sqrt{N}\,)$; if $k_{{\rm max}}$ is taken any larger,  additional correlations occur between high degree vertices \cite{PhysRevE.71.027103}, and the failure rate becomes unacceptable. Alternatively, we could have larger values of $k_{{\rm max}}$ by increasing the graph size, but that would make the numerical diagonalisation computationally expensive. The degree sequences are generated randomly and checked for graphicality before generating the graph.
\begin{figure}
\psfrag{y}[B][B][1][-90]{$\rho(\lambda)\quad$}\psfrag{x}{$\lambda$}
\includegraphics[trim=0 50 0 50, width=\textwidth]{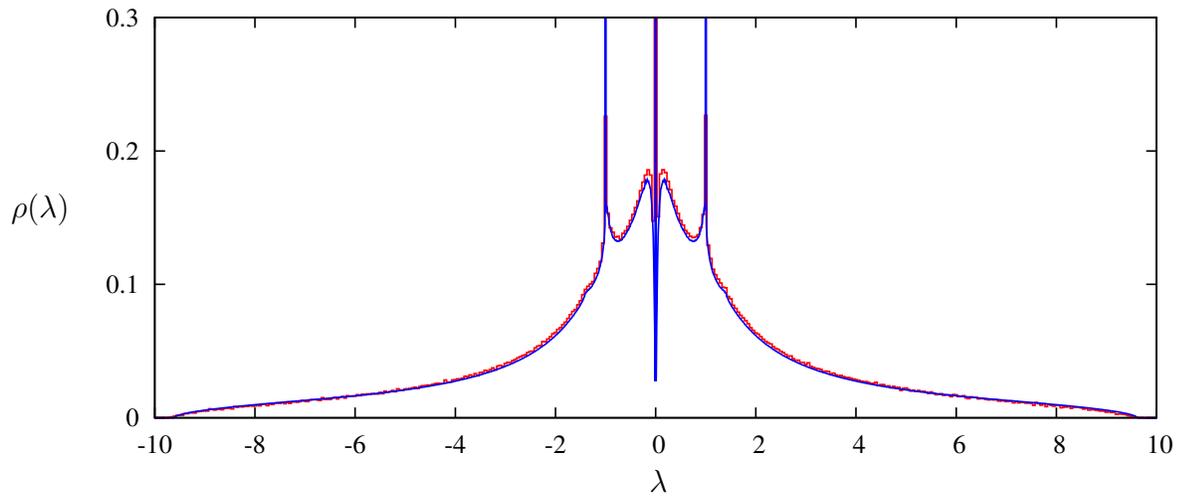}
\caption{Comparison of the results of population dynamics (blue line) and direct diagonalisation (red histogram) for the choice $P(k,k')\propto\tau^{kk'}$, where $\tau=0.999$ and $k_{\max}=45$. The adapted Steger-Wormald algorithm was used to generate 500 graphs of size $N=2000$, whose eigenvalues were used to construct the histogram.}
\label{sf}
\end{figure}\\
The spectral density can again be computed via population dynamics. Figure \ref{sf} shows a comparison between the results of population dynamics and a histogram of eigenvalues from 500 random graphs of size $N=2000$. In this case we take a population of $\mathcal{N}=10^3$ variables $\{\Delta_{i}(k)\}_{i=1}^{\mathcal{N}}$, which are iterated over 100 MC steps. The spectral density is then averaged over a further 50 MC steps.\\ 
As before some of the salient features of the spectral density can be intuitively explained in terms of the underlying graph structure. First we note that the spectral density for $k_{\max}=45$ and $\tau=0.999$ has a bounded support as expected due to  the Perron-Frobenius theorem, but as we take the $k_{\max}\to\infty$ and $\tau\to1$,  the mean degree diverges and the spectral density presents  heavy tails. It is interesting to analyse the contribution of vertices of high degree to the spectral density. As mentioned in \cite{Dorogovtsev2003}, for such vertices it may be sufficient to consider only the mean behaviour of the neighbouring vertices. In this  effective medium approximation (EMA) the approximate behaviour of the tails for very large $|\lambda|$ reads $\rho(\lambda)\simeq 2k_\lambda p(k_\lambda)/|\lambda|$, where $k_\lambda=\lambda^2+\mathcal{O}(1)$. A more rigorous analysis in \cite{Mihail2002} states that the largest eigenvalues of graphs with heavy-tailed degree distributions occur close to the square roots of the largest degrees.\\
In Figure \ref{tail} we show the results of population dynamics simulations in the tail of the spectral density for a much larger maximum degree of $k_{\max}=400$. The approximate curve given by the EMA gives a reasonable fit with the result of the simulation and, as expected, the density drops dramatically shortly after $\sqrt{k_{\max}}=20$. The contributions to the density coming from high degree vertices can be isolated in the output of the population dynamics algorithm; we have included in Figure \ref{tail} contributions from several high degrees $k$, each of which exhibits sharp peak close to $\sqrt{k}$.
\\
The other main feature of the density shown in Figure \ref{sf} is the presence of Dirac delta peaks at -1, 0 and 1 whose weight may be bounded by using   the distributions $p(k)$ and $P(k|k')$. For instance, the weight to the peaks at $\pm1$ has contributions from connected pair of vertices of degree 1, which for the choice $k_{\max}=45$ and $\tau=0.999$ have a likelihood of $p(1)P(1|1)/2=0.0031$, not far from the exact weight of $0.0037$  obtained from numerical diagonalisation. A similar intuitive argument can be used to obtain a bound for the weight of the Dirac delta peak at zero, whose appearance is due to dead-end vertices \cite{Dorogovtsev2003,Golinelli2003}.
\subsection{Case $L=2$. Levels of approximation}
\begin{figure}
\centering\includegraphics[trim=200 200 170 00, width=\textwidth]{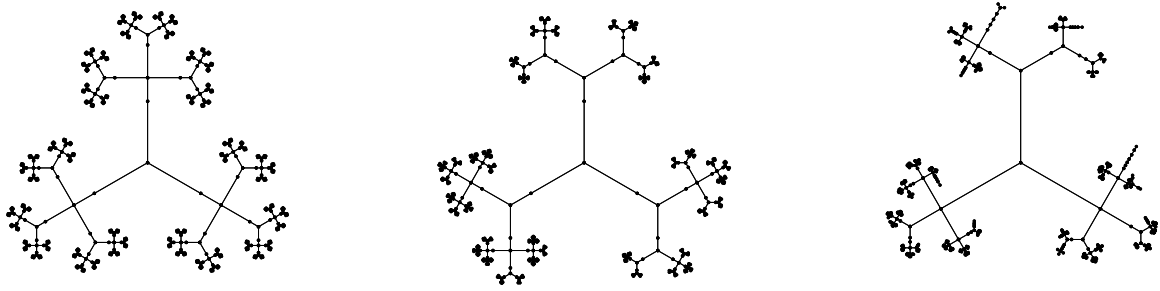}
\caption{Left - a typical neighbourhood of a vertex of degree 3  of a random graph specified by (\ref{condk}). Middle - a neighbourhood of a vertex of degree 3 in a random graph with degree distribution \eref{pknice} and $P(k|k')$ given by \eref{Pkknice}. Right - a neighbourhood of a vertex of degree 3 in an uncorrelated random graph with degree distribution \eref{pknice}.}
\label{graph}
\end{figure}

\begin{figure}
\psfrag{y}[B][B][1][-90]{$\rho(\lambda)\quad$}\psfrag{x}{$\lambda$}
\centering\subfloat[Comparison of the exact spectral density given by equation \eref{rho32} (blue line) and direct diagonalisation (red histogram). We have checked the isolated peaks at $\lambda\simeq \pm\sqrt{7}$ are due to finite size effects.]{\includegraphics[trim=0 60 0 60, width=0.86\textwidth]{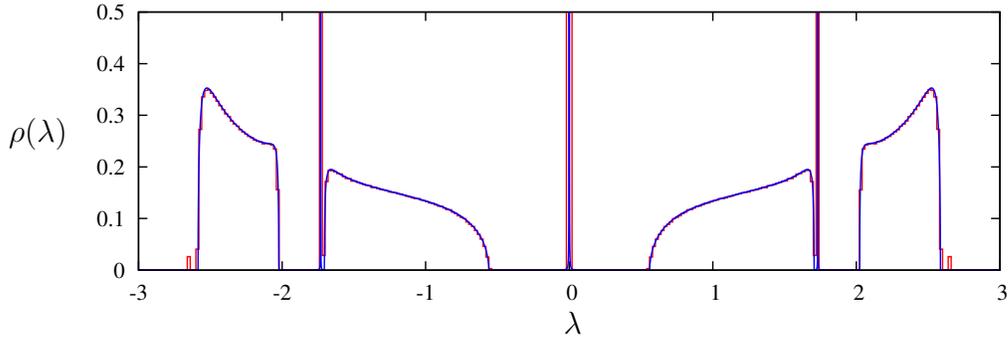}}\\
\subfloat[Comparison of population dynamics with degree distribution \eref{pknice} and degree-degree correlations \eref{Pkknice} (blue line) and direct diagonalisation (red histogram). To visualise the Dirac delta peak we have taken $\varepsilon=10^{-6}$.]{\includegraphics[trim=0 60 0 60, width=0.86\textwidth]{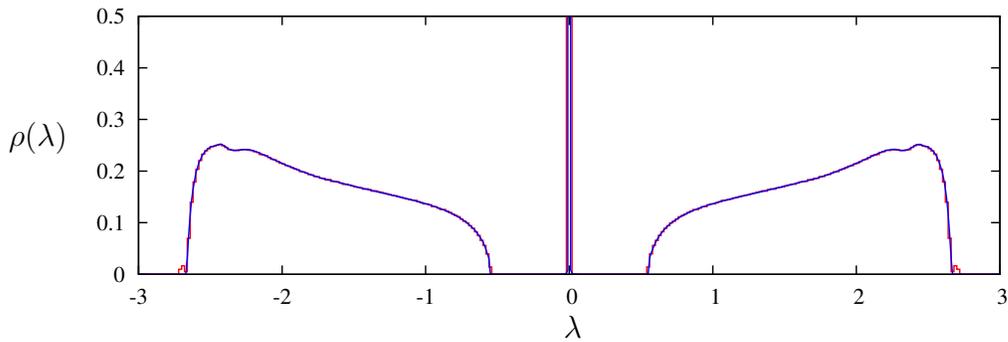}}\\
\subfloat[Comparison of population dynamics with degree distribution only (blue line) and direct diagonalisation (red histogram).]{\includegraphics[trim=0 60 0 60, width=0.86\textwidth]{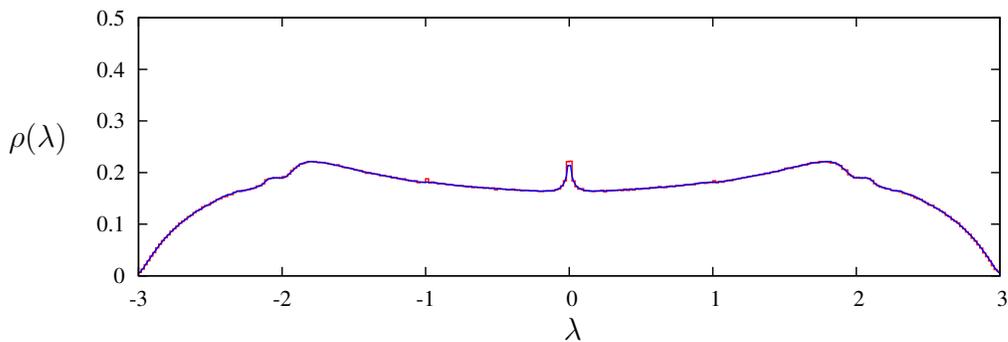}}\\
\caption{\label{levels}Comparison theoretical results (blue lines) and direct diagonalisation (red histograms) for the different levels of approximation to the ensemble specified by (\ref{condk}). In plot (a), the blue line shows the exact result for the spectral density, given by \eref{rho32}, in plots (b) and (c) the result of population dynamics is shown. We use the method outlined in the text to generate graphs for plot (a), and the adapted Steger-Wormald algorithm for (b) and (c). In each case, 1000 graphs of size $N=1900$ were generated and diagonalised.}
\end{figure}

Suppose that the exact knowledge of a graph ensemble is reduced solely to a set of statistical properties captured by, for instance, the degree distribution $p(k)$ and the conditional distribution $P(k|k')$. Whilst in a few cases such quantities suffice to fully characterise the graph ensemble, this is not generally true. We would like to understand in which way the lack of more accurate information affects the spectral density.\\
With this in mind let us consider a graph ensemble with generalised degrees $\{\bk_i\}_{i=1}^N$ such that $\bk_i\in \{\kappa_2,\kappa_3,\kappa_4\}$, where
\begin{equation}
\kappa_2=\left(\begin{array}{c}2\\7\end{array}\right)\,,\quad\kappa_3=\left(\begin{array}{c}3\\6\end{array}\right)\,,\quad\kappa_4=\left(\begin{array}{c}4\\8\end{array}\right) \,.
\label{condk}
\end{equation}
This ensemble is composed of graphs with vertices of degrees 2, 3 and 4. Moreover, those vertices of degree 2 must be connected to one vertex of degree 3 and one of degree 4 and those of degrees 3 and 4 must be connected to vertices of degree 3 only. A portion of such graph is shown in the leftmost part of Figure \ref{graph}.
\\
For this graph ensemble a quick counting argument gives the following expressions for $p(k)$ and $P(k|k')$
\numparts
\begin{eqnarray}
&p(k)=\frac{12}{19}\delta_{k,2}+\frac{4}{19}\delta_{k,3}+\frac{3}{19}\delta_{k,4}\,,\label{pknice}\\
&P(k|k')=\delta_{k,2}\delta_{k',4}+\delta_{k,2}\delta_{k',3}+\left(\frac{1}{2}\delta_{k,3}+\frac{1}{2}\delta_{k,4}\right)\delta_{k',2}\label{Pkknice}\,,
\end{eqnarray}
\endnumparts
knowledge of which does not fully characterise the graph ensemble. Note that this example is such that the set of self-consistency equations \eref{taufinal} can be solved exactly. To do so we first need the conditional distribution $P\left(\{\bi{q}_t\}_{t=1}^{k-1}|\bq,\bi{k}\right)$ which in this case reads
\begin{eqnarray}
&P(\kappa_3|\kappa_4,\kappa_2)=1\,,\quad P(\kappa_4|\kappa_3,\kappa_2)=1\,,\nonumber\\
&P(\kappa_2,\kappa_2|\kappa_2,\kappa_3)=1\,,\quad P(\kappa_2,\kappa_2,\kappa_2|\kappa_2,\kappa_4)=1\,,\nonumber
\end{eqnarray}
or zero otherwise.  This results in a set of self-consistency equations for the densities $\{\psi(\Delta|\kappa_3,\kappa_2),\psi(\Delta|\kappa_2,\kappa_3),\psi(\Delta|\kappa_4,\kappa_2),\psi(\Delta|\kappa_2,\kappa_4)\}$ that admits a solution of the type $\psi(\Delta|\kappa_a,\kappa_b)=\delta(\Delta-\Delta_{a,b})$ with $\Delta_{a,b}$ obeying a simple set of algebraic equations
\begin{eqnarray}
\fl\Delta_{3,2}=\frac{1}{i\lep+\Delta_{2,4}}\,,\,\,
\Delta_{4,2}=\frac{1}{i\lep+\Delta_{2,3}}\,,\,\,
\Delta_{2,3}=\frac{1}{i\lep+2\Delta_{3,2}}\,,\,\,
\Delta_{2,4}=\frac{1}{i\lep+3\Delta_{4,2}}\,.
\label{niceD}
\end{eqnarray}
Finally, to find an expression for the spectral density we first need the distribution $P(\{\bi{q}_t\}_{t=1}^{k}|\bi{k})$ which in this case reads
\begin{eqnarray}
&P(\kappa_3,\kappa_4|\kappa_2)=\frac{1}{2}\,,\quad P(\kappa_4,\kappa_3|\kappa_2)=\frac{1}{2}\,,\nonumber\\
&P(\kappa_2,\kappa_2,\kappa_2|\kappa_3)=1\,,\quad P(\kappa_2,\kappa_2,\kappa_2,\kappa_2|\kappa_4)=1\,,\nonumber
\end{eqnarray}
or zero otherwise. This yields
\begin{equation}
\hspace{-1cm}\rho(\lambda)=\lim_{\varepsilon\to0^{+}}\frac{1}{19 \pi}\textrm{Re}\left[\frac{12}{i\lep+\Delta_{2,3}+\Delta_{2,4}}+\frac{4}{i\lep+3\Delta_{3,2}}+\frac{3}{i\lep+4\Delta_{4,2}}\right]\,.
\label{rhonice}
\end{equation}
Upon solving \eref{niceD}, plugging the solutions into \eref{rhonice} and carefully analysing the poles, we can write
\begin{eqnarray}
\rho(\lambda)&=\frac{5}{19}\delta(\lambda)+\frac{1}{19}\delta(\lambda+\sqrt{3})+\frac{1}{19}\delta(\lambda-\sqrt{3})\nonumber\\
&+\frac{12| 2 \lambda^2-7|\sqrt{-25 - \lambda^2 (-7 + \lambda^2) (14 - 7 \lambda^2 + \lambda^4)} }{19\pi|\lambda(\lambda^2-4)(\lambda^2-7) (\lambda^2-3)|}\mathbb{I}_{{\mathcal R}}(\lambda)
\label{rho32}
\end{eqnarray}
with ${\mathcal R}=[\lambda^{-}_{+},\lambda^{-}_{-}]\cup[\lambda^{+}_{-},\lambda^{+}_{+}]$ and
\begin{eqnarray}
\lambda^{\mu}_\sigma=\frac{1}{2}\sqrt{14 +2\mu \sqrt{21 + 8\sigma \sqrt{6}}}
\end{eqnarray}
with $\sigma,\mu\in\{-,+\}$ and where $\mathbb{I}_{{\mathcal R}}(\lambda)=1$ if $|\lambda|\in{\mathcal R}$ or zero otherwise.\\
This example was chosen specifically to keep the local structure of the graphs deterministic and thus make equations \eref{taufinal} exactly solvable in the case $L=2$. If we reduce the statistical information we have about the ensemble, either by taking $L=1$, or assuming the degrees to be uncorrelated, the local graph structure becomes random, and the resulting expressions are no longer exactly solvable. Figure \ref{graph} shows the randomising effect of these simplifications.\\
We have used population dynamics to compute the spectral density of random graphs with degree distribution \eref{pknice} and degree-degree correlations \eref{Pkknice}, as well as uncorrelated random graphs with degree distribution \eref{pknice}. For each degree, the population size is $\mathcal{N}=10^4$, iterated over 200 MC steps. The spectral density is then averaged over 50 MC steps. The results are shown in Figures \ref{levels}(b) and \ref{levels}(c), alongside histograms of the eigenvalues of 1000 graphs of size $N=1900$, generated using the adapted Steger-Wormald algorithm \footnote{It is necessary to slightly modify the adapted Steger-Wormald algorithm to make sure the constraints specified by (\ref{condk}) are met.}. Figure \ref{levels}(a) shows the exact spectral density given by \eref{rho32}, plotted alongside a histogram of eigenvalues obtained from randomly generated graphs of that type.\\ 
In Figure \ref{levels}, the effects of reducing knowledge of an ensemble (and hence increasing randomness) are clearly visible. In addition to a general smoothing effect, which one might expect, the most striking feature is the appearance of gaps in the spectral density. When the ensemble is fully specified by (\ref{condk}), the continuous part of the spectral density is divided into four disjoint components. When one specifies only the degree distribution and degree-degree correlations, the number of components reduces to two, and when only the degree distribution is known, there is no gap in the density at all.\\ 
The appearance of the gap can be traced back to the periodicity in the random graph ensembles (see \cite{Viswanath1994}). For instance, in the original ensemble a  walker moving away from a central vertex will repeatedly visit vertices of degree sequence $\{\ldots,2,4,2,3,2,4,2,3,\ldots\}$ of periodicity four which is also manifestly explicit in the equations \eref{niceD}. In the case that the degree-degree correlations are specified, whilst the degree sequence of a walk is now random and therefore not strictly periodic, it is still true that every other vertex visited will have degree 2 (see Figure \ref{graph}), periodicity which is enough to split the spectral density into two components. In the last case, where only the degree distribution is known, the sequence is fully random (see Figure \ref{graph}), and the resulting spectral density has no gap.
\subsection{Community structure}
\begin{figure}
\psfrag{y}[B][B][1][-90]{$\rho(\lambda)\quad$}\psfrag{x}{$\lambda$}
\includegraphics[trim=0 50 0 50, width=\textwidth]{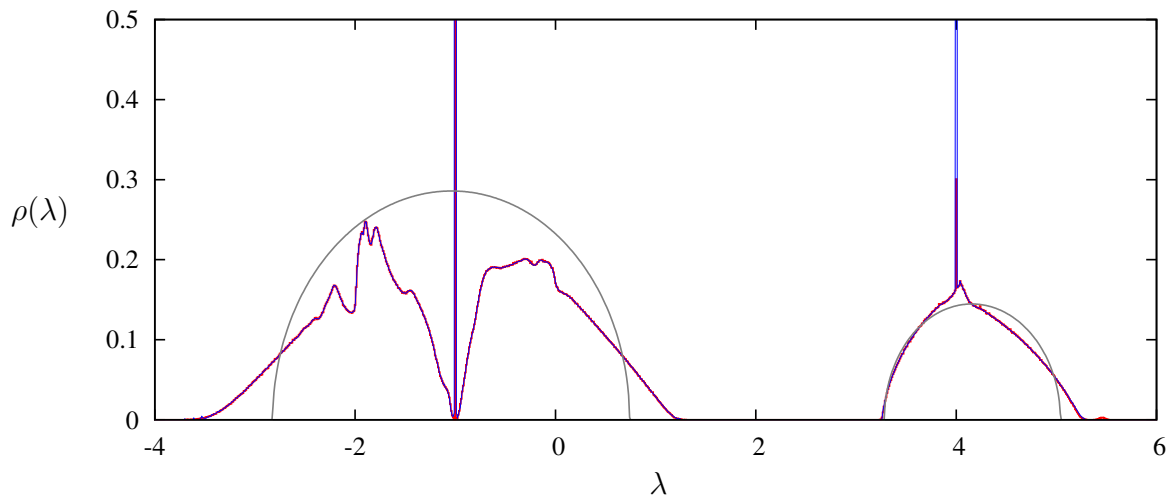}
\caption{Comparison of the results of population dynamics (blue line) and direct diagonalisation (red histogram) for the community structure ensemble described in the text. The grey curve shows the high connectivity limit for this ensemble.}
\label{comm}
\end{figure}

The population dynamics algorithm used to solve \eref{taufinal} is easily adapted to solve the equivalent self-consistency equation \eref{taucomm} for the communities model; one simply initialises populations of $M\times M$ matrices $\bi{\Delta}$ and updates them according to \eref{taucomm}.\\ 
The presence of communities in a graph typically results in a very different spectral density, to illustrate this we consider a simple choice for the community structure ensemble. Suppose we have communities given by the complete graph on $M$ vertices, connected in a Poissonian random graph of average degree $c$, in which connected communities are joined by a single randomly drawn edge. In the weight \eref{defbWclus}, this corresponds to the choices $Q(\bk_i,\bk_j)=1$ for all $i,j$, $\nu(A)=\big(\prod_{a}\delta_{A_{aa},0}\big)\big(\prod_{a\neq b}\delta_{A_{ab},1}\big)$, and $\mu(B)=M^{-2}\sum_{a,b}\delta_{B_{ab},1}\prod_{(c,d)\neq(a,b)}\delta_{B_{cd},0}\,$.\\
Taking $M=5$ and $c=5$, we use population dynamics to solve \eref{taucomm} for this ensemble, the spectral density is them computed using \eref{rhocomm}. To compare with the results of direct diagonalisation, 1000 graphs of size $N=5000$ were generated. A histogram of their eigenvalues, alongside the result of population dynamics, is shown in Figure \ref{comm}.
\\
It is well known that in the limit $c\to\infty$ the spectral density of a Poissonian random graph with average degree $c$, and edges of weight $1/\sqrt{c}$, converges to Wigner's semi-circular distribution \cite{Rodgers1988, Rogers2008}. We can compute a generalisation of this result for the community ensemble considered here through an appropriate treatment of the self-consistency equation \eref{taucomm}. It is necessary to re-weight the edges between communities, in order to keep the spectral density bounded as $c\to\infty$, we take $\mu(B)=M^{-2}\sum_{a,b}\delta_{B_{ab},\sqrt{M/c}}\prod_{(c,d)\neq(a,b)}\delta_{B_{cd},0}\,$. Keeping only the terms relevant in the $c\to\infty$ limit we obtain an expression for the mean $\bi{\Delta}$,
\begin{equation}
\langle \bi{\Delta} \rangle =\Big(i(\lep I_M-K_M)+c\Big\langle B\langle \bi{\Delta}\rangle B^T\Big\rangle_B\Big)^{-1}\,,
\end{equation}
where $K_M$ is the connectivity matrix of the complete graph on $M$ vertices. For the above choice of $\mu(B)$, we have $\langle B\bi{\Delta}B^T\rangle=(1/cM)\Delta I_M$, where $\Delta=(1/M)\Tr\langle \bi{\Delta} \rangle$. Diagonalising $K_M$, we obtain a cubic equation for $\Delta$,
\begin{equation}
\Delta=\left(\frac{M-1}{M}\right)\frac{1}{i\lep+\Delta+i}+\left(\frac{1}{M}\right)\frac{1}{i\lep+\Delta-(M-1)i}\,.
\label{cubic}
\end{equation}
In the case $M=5$, we solve \eref{cubic} to find the following expression:
\begin{equation}
\rho(\lambda)=\frac{\sqrt{3}}{2\pi}\left|u-\frac{\lambda^2-3\lambda+18}{9u}\right|\mathbb{I}_\mathcal{D}(\lambda)\,,
\end{equation}
where 
\begin{equation}
u=\left|\,\frac{1}{27}\lambda^3-\frac{1}{6}\lambda^2-2\lambda+4+\frac{1}{18}\sqrt{3d}\,\right|^{1/3},
\end{equation}
and $d=-25\lambda^4+156\lambda^3+72\lambda^2-1296\lambda+864$. The domain is given by $\mathbb{I}_\mathcal{D}(\lambda)=1$ if $d>0$, and zero otherwise.

 \section{Conclusions}
Past work on the spectral density of random graphs has typically been confined to simple ensembles in which at most the degree distribution is specified and even then exact results have only been obtained relatively recently. At the same time, the field of complex networks has gained a great deal of attention from all over the scientific community. Before the analysis of spectral density can be used to provide insights into the behaviour and characterisation of complex networks, progress must first be made to expand the class of ensembles for which exact results are obtainable.\\
In this paper we have sought to do just this, through the calculation of the spectral density of random graphs with constrained topologies. Complex correlations between the degrees of non-neighbouring vertices are incorporated in the constrained generalised degree ensemble and we also introduce a simple extension of this model to one featuring a community structure. The important statistical properties of the constrained degree ensemble are captured in the distribution \eref{PKN}, which we compute via a saddle point analysis in the large $N$ limit. This calculation foreshadows the replica calculation of the spectral density and provides important insight.\\
For the problem of determining the mean spectral density, we take standard steps to map the problem onto one of an interacting system of  dynamical variables, to which the replica method is applied. Following recent advances, the form of replica symmetric ansatz is identified as a superposition of Gaussians. Exploiting the insights gained in the earlier calculation, we obtain closed expressions for the spectral density in terms of the statistical properties of the graph ensemble \eref{taufinal}. Similar equations are found for the community structure ensemble \eref{taucomm}.\\
Though the resulting equations may not often have easily found analytic solutions, they can be efficiently solved numerically using the population dynamics algorithm we describe earlier. In this way it is possible to analyse the spectral density of a given graph ensemble without the need to generate and diagonalise large numbers of graphs. An instance of this is provided by the discussion of the tails in the example with a power-law degree distribution; here the results of population dynamics could not feasibly be obtained by diagonalising random matrices (for $k_{\max}=400$, we would require graphs of around $1.6\times 10^{5}$ vertices). We hope that the methods discussed here will prove useful tools in the study of complex networks. \\
Although, as we have demonstrated, the statistics of generalised degrees can have a significant impact on the spectral density of the graph ensemble, this is certainly not the only factor at work. One aspect of the topology of complex networks which may play an important role, but which we have not considered so far, is the statistics of the loops in the graph. Unfortunately, knowledge of the generalised degrees of a graph gives no information about loops, and hence the constrained generalised degree ensemble is not likely to be useful to study of their effect on spectral density. Looking to the future, it seems the next major step forward in the analysis of spectral density of random graphs will require techniques capable of handling the effects of loops. Several new techniques to correct for the presence of loops in related problems have a appeared recently \cite{Montanari2005,Parisi2006,Chertkov2006}, and we hope that similar ideas may also be applied to the study of spectral density.

\ack

K.T. acknowledges hospitality from the Disordered Systems Group,   at the department of Mathematics, King's College London.    K.T. is supported by a Grant-in-Aid Scientific Research on Priority Areas `Deepening and Expansion of Statistical  Mechanical Informatics (DEX-SMI)' from MEXT, Japan (No. 18079006). Conrad P\'{e}rez Vicente acknowledges funding from `Ministerio de Educaci\'on y Ciencia' FIS2006-13321-C02-01.

\newpage
\bibliographystyle{iopart-num}
\bibliography{constrained}

\providecommand{\newblock}{}
\begin{thebibliography}{10}
\expandafter\ifx\csname url\endcsname\relax
  \def\url#1{{\tt #1}}\fi
\expandafter\ifx\csname urlprefix\endcsname\relax\def\urlprefix{URL }\fi
\providecommand{\eprint}[2][]{\url{#2}}

\bibitem{Guhr1998}
Guhr T, M\"uller-Groeling A and Weidenm\"uller H~A 1998 {\em Phys. Rep.\/} {\bf
  299} 190

\bibitem{Wigner1958}
Wigner E~P 1958 {\em Ann. Math.\/} {\bf 67} 325

\bibitem{Dyson1962}
Dyson F~J 1962 {\em J. Math. Phys.\/} {\bf 3} 140

\bibitem{Mehta1991}
Mehta M~L 1991 {\em Random Matrices\/} (New York: Academic Press)

\bibitem{Rodgers1988}
Rodgers G~J and Bray A~J 1988 {\em Phys. Rev. B\/} {\bf 37} 3557

\bibitem{Bauer2001}
Bauer M and Golinelli O 2001 {\em J. Stat. Phys.\/} {\bf 103} 301

\bibitem{Biroli1999}
Biroli G and Monasson R 1999 {\em J. Phys. A\/} {\bf 32} L255

\bibitem{Dorogovtsev2003}
Dorogovtsev S~N, Goltsev A~V, Mendes J~F~F and Samukhin A~N 2003 {\em Phys.
  Rev. E\/} {\bf 68} 046109

\bibitem{Mirlin1991}
Mirlin A~D and Fyodorov Y~V 1991 {\em J. Phys. A\/} {\bf 24} 2273

\bibitem{Nagao2007}
Nagao T and Tanaka T 2007 {\em J. Phys. A\/} {\bf 40} 4973

\bibitem{Nagao2008}
Nagao T and Rodgers G~J 2008 {\em J. Phys. A\/} {\bf 41} 265002

\bibitem{Semerjian2002}
Semerjian G and Cugliandolo L~F 2002 {\em J. Phys. A\/} {\bf 35} 4837

\bibitem{Rogers2008}
Rogers T, {P\'erez Castillo} I, K\"uhn R and Takeda K 2008 {\em Phys. Rev.
  E.\/} {\bf 78} 031116

\bibitem{Kuhn2008}
K\"uhn R 2008 {\em J. Phys. A\/} {\bf 41} 295002

\bibitem{Dean2002}
Dean D~S 2002 {\em J. Phys. A\/} {\bf 35} L153

\bibitem{Ciliberti2005}
Ciliberti S, Grigera T~S, Mart{\'{\i}}n-Mayor V, Parisi G and Verrocchio P 2005
  {\em Phys. Rev. B\/} {\bf 71} 153104

\bibitem{Bordenave2007}
Bordenave C and Lelarge M 2007 {\em Eprint arXiv.org:0801.0155\/}

\bibitem{10}
Fortunato S 2009 {\em Eprint arXiv.org:0906.0612\/}

\bibitem{Farkas2001}
Farkas I~J, Derenyi I, Barabasi A~L and Vicsek T 2001 {\em Phys. Rev. E\/} {\bf
  64} 026704

\bibitem{Nikos}
Skantzos N {\em Unpublished report\/}

\bibitem{Coolen2008}
Bianconi G, Coolen A~C~C and {P\'erez Vicente} C~J 2008 {\em Phys. Rev. E\/}
  {\bf 78} 016114

\bibitem{Rogers2009}
Rogers T and {P\'erez Castillo} I 2009 {\em Phys. Rev. E\/} {\bf 79} 012101

\bibitem{Vicente2008}
{P\'erez Vicente} C~J and Coolen A~C~C 2008 {\em J. Phys. A\/} {\bf 41} 255003

\bibitem{Oraby2007}
Oraby T 2007 {\em J. Theor. Prob.\/} {\bf 20} 1572

\bibitem{Jalan2008}
Jalan S 2009 {\em Phys. Rev. E\/} {\bf 80} 04610

\bibitem{Reimer2009}
Ergun G and K\"uhn R 2009 {\em Eprint arXiv.org:0908.3155\/}

\bibitem{Edwards1976}
Edwards S~F and Jones R~C 1976 {\em J. Phys. A\/} {\bf 9} 1595

\bibitem{dorogovtsev2008}
Dorogovtsev S~N, Goltsev A~V and Mendes J~F~F 2008 {\em Rev. Mod. Phys.\/} {\bf
  80} 1275

\bibitem{Berlin1952}
Berlin T~H and Kac M 1952 {\em Phys. Rev.\/} {\bf 86} 821

\bibitem{Mezard2001}
M\'ezard M and Parisi G 2001 {\em Eur. Phys. Jour. B\/} {\bf 20} 217

\bibitem{Steger1999}
Steger A and Wormald N~C 1999 {\em Comb. Prob. Comp.\/} {\bf 8} 377

\bibitem{1459763}
Bayati M, Kim J~H and Saberi A 2007 {\em Proceedings of the 10th International
  Workshop on Approximation and the 11th International Workshop on
  Randomization, and Combinatorial Optimization. Algorithms and Techniques\/}
  (Springer-Verlag) p 326

\bibitem{Blitzstein2006}
Blitzstein J and Diaconis P 2006 {\em Unpublished manuscript\/}

\bibitem{PhysRevE.71.027103}
Catanzaro M, Bogu\~n\'a M and Pastor-Satorras R 2005 {\em Phys. Rev. E\/} {\bf
  71} 027103

\bibitem{Mihail2002}
Mihail M and Papadimitriou C 2002 {\em Lect. Notes Comput. Sci.\/} {\bf 254}
  2483

\bibitem{Golinelli2003}
Golinelli O 2003 {\em Eprint cond-mat/0301437\/}

\bibitem{Viswanath1994}
Viswanath V~S and M\"uller G 1994 {\em The Recursion Method: Application to
  Many-Body Dynamics\/} (Berlin: Springer)

\bibitem{Montanari2005}
Montanari A and Rizzo T 2005 {\em J. Stat. Mech. : Theor. Exp.\/}  P10011

\bibitem{Parisi2006}
Parisi G and Slanina F 2006 {\em J. Stat. Mech. : Theor. Exp.\/}  L02003

\bibitem{Chertkov2006}
Chertkov M and Chernyak V~Y 2006 {\em J. Stat. Mech. : Theor. Exp.\/}  P06009

\end{thebibliography}
\end{document}